\documentclass[12pt]{iopart}
\expandafter\let\csname equation*\endcsname\relax
\expandafter\let\csname endequation*\endcsname\relax
\usepackage{amsmath}
\usepackage{amssymb}
\usepackage{amsthm}
\usepackage{mathtools}
\usepackage{bm}
\usepackage{calc}
\usepackage{graphicx}
\usepackage[caption=false]{subfig}
\usepackage[usenames]{color}

\usepackage{ulem}

\renewcommand{\vec}[1]{{\bm{#1}}}

\begin{document}

\title[Adsorption of annealed branched polymers on curved surfaces]{Adsorption of annealed branched polymers on curved surfaces}

\author{Jef Wagner, Gonca Erdemci-Tandogan, Roya Zandi } 
\address{ Department of Physics and Astronomy, University of California, Riverside, CA 92521, USA}


\begin{abstract}
  The behavior of annealed branched polymers near adsorbing surfaces
  plays a fundamental role in many biological and industrial processes. Most importantly single stranded RNA in solution tends to fold up and
self-bind to form a highly branched structure. Using a mean field theory, we both perturbatively and numerically examine the adsorption of branched polymers on surfaces of several different geometries in a good solvent. Independent of the geometry of the wall, we observe that as branching density increases, surface tension decreases. However, we find a coupling between the branching density and curvature in that a further lowering of surface tension occurs when the wall curves towards the polymer, but the amount of lowering of surface tension decreases when the wall curves away from the polymer. We find that for branched polymers confined into spherical cavities, most of branch-points are located in the vicinity of the interior wall and the surface tension is minimized for a critical cavity radius. For branch polymers next to sinusoidal surfaces, we find that branch-points accumulate at the valleys while end-points on the peaks. 
\end{abstract}

\section{Introduction}

Branched polymers play important roles in many biological and
industrial systems, notable among them single stranded RNA (ssRNA) that in
solution takes on a branched secondary structure \cite{deGennes:68a, Grosberg:95a, Brion:97a, Gutell:85a, Daoud:90a, Nguyen:06a, Lee:08a,Paul:13a,Higgs,Hagan}. Recent experiments on viruses show that some viral RNAs, in particular, assume highly branched structures \cite{Gopal,Garmann}. The physics of
polymer adsorption on different kinds of interfaces has,
specifically, attracted a lot of interest for over half a
century \cite{deGennes:69a, deGennes:81a, deGennes:82a, deGennes:83a, Jones:77a, Fleer:82a, Pincus:84a, Ober:83a, Eisenriegler:77a, diMeglio:83a, Ji:87a, Winkler}.  In particular, it has been shown that polymer topology can effect the thermodynamic behavior of polymers near surfaces \cite{Carignano1994}. More recently the adsorption of RNA onto spherical gold
nano-particles has been the focus of intense research because of its
potential application in drug delivery or gene therapy \cite{Sun:07a, Delong:10a,Johnson:12a,Rothberg:05a,ChenDragnea2006,SiberZandi2010, Ding2014, Cai2014,Tiwari2011}. 

RNA is
considered as an {\it annealed} branched polymer mainly due to the
fact that the base-pair binding in RNA is often weak enough that the
branching structure can change due to thermal fluctuations \cite{Grosberg:95a,Grosberg:97a}. Beyond
the adsorption of RNA onto nano-particles, the behavior of annealed
branched polymers next to surfaces of complex geometries is
intriguing. Despite the presence of several excellent books and
review articles, the impact of branching on adsorption of biopolymers
at planar or rough substrates is yet not well-studied.

Several experiments compare the efficiency--directly connected to the free energy--of encapsidation of linear polymers and viral RNAs by virus capsid proteins\cite{Comas,Chuck2008,Cadena2011}. Field theoretic models have been used extensively to calculate the free energy of linear
polymers \cite{Gaspari:86a, Joanny:99a, Borukhov:98a, Ji:88a, Vanderschoot2009,Vanderschoot2007,Siber:08a,Zandi:03a,Zandi:01a}. In a 1972 seminal paper de Gennes noted an equivalence between the statistics of a self-avoiding polymer and the $n\to0$ limit of an $\mathcal{O}(n)$ model of a magnet \cite{deGennes:72a}, see Appendix A for a review of $\mathcal{O}(n)$ model. Using this observation,  which relates a mathematically interesting but unphysical limit for the model of a magnet to the statistics of polymers,
it became possible to use the tools of statistical field theory to describe the
physical properties of a polymer solution.  Later, de Gennes field theory was expanded to describe the statistics of
{\it annealed} branched polymers \cite{Lubensky:79a, Elleuch:95a}. 


In this paper we use a mean field theory to study the adsorption of
annealed branched polymers on different types of surfaces from a semi-dilute polymer solution in a good solvent.  We study the effect of curvature by examining the adsorption onto the exterior and interior of a spherical surface, and investigate the impact of roughness by examining the adsorption onto a sinusoidal surface. Instead of considering a random roughness, we employ a grating geometry because of its enormous mathematical simplifications and the fact that it has been shown that qualitatively the essential features of the results are the same \cite{Ji:87a, Ji:88a}. 

By numerically solving the relevant nonlinear equations we find that compared to the adsorption to a flat wall, branching density, surface tension, and the monomer density all increase if the polymer is adsorbed onto the interior wall of a spherical cavity but decrease if adsorbed on the exterior surface of the sphere. While our results show that surface tension always decreases as branching density increases independent of the geometry of the wall, we find the interplay of curvature and branching density conspires to further lower the surface tension when the wall curves toward the polymers but lessens the amount of decrease in the surface tension when the wall curves away from the polymer. In the limit of large spheres, we solve the nonlinear equations perturbatively, which match the numerical results.  Furthermore, in case of sinusoidal surfaces, we find inhomogeneity in the branching density as it increases in the valleys but decreases in the peaks. 

The remainder of this paper is organized as follows.  In the first section
we present our mean field approach and in the following section we
use this method to investigate the impact of branching combined with
surface curvature on polymer adsorption. In particular we will
examine what effect the branching structure has on the adsorption of
polymers to nano-spheres and sinusoidal surfaces.  We will finish with a brief summary and present our main conclusions. 

In the appendix, for completeness and pedagogical reasons, we derive a simple field theoretical model for a
branched polymer by revisiting the field
theory developed by Isaacson and Lubensky\cite{Lubensky:79a} for
branched polymers and will spell out in detail the equivalence
between the polymer statistics and the $n\to0$ limit of the
$\mathcal{O}(n)$ model.

\section{Mean Field Approximation}\label{sec:MF} 
To describe a branched polymer on a lattice, we assume the polymer
system consists of branch-points and end-points lying on the lattice
sites, and bonds that join neighboring lattice sites. We treat the
system as an annealed branched polymer, so the structure of the
branched polymer is not fixed. For simplicity, we assume that all
branch-points are exactly of order three because all higher order branch-points can be considered as many order three branch-points in
close proximity to each other. For example, two order three branch-points close together will behave very similarly to an order four
branch-point. The quantities that describe such a polymer system are:
(i) $N_p$, the number of polymers; (ii) $N_b$, the number of bonds;
(iii) $N_1$, the number of endpoints; (iv) $N_3$, the number of
branch-points; and (v) $N_l$, the number of loops.  There is a
constraint \cite{Lubensky:79a} relating most of these quantities such that
\begin{equation}\label{eq:noloop}
2 (N_p - N_l)= N_1 - N_3.
\end{equation}
In this paper, we consider a system of branched polymers with no loops and set
$N_l=0$. 

The primary statistical quantity of interest is the multiplicity
$\Omega(N_b, N_1, N_3; V)$, defined as the number of ways to arrange a polymer
system of $N_b$ bonds, $N_1$ end-points, and $N_3$ branch-points on a lattice
that occupies a volume $V$. This quantity is equivalent to the number of microstates for the microcanonical ensemble. From the multiplicty, we can form the grand canonical partition function
\begin{multline}\label{eq:grand_partition}
\Xi(K,f_1,f_3; V) = \\ \sum_{\mathclap{N_b,N_1,N_3}}
K^{N_b} f_1^{N_1} f_3^{N_3} \Omega(N_b, N_1, N_3; V),
\end{multline}
where $K$, $f_1$, and $f_3$ are the fugacities for the bonds,
end-points, and branch-points respectively. Note that we use Eq.~\eqref{eq:noloop} along with the assumption that $N_l=0$
to eliminate the dependence upon $N_p$. From this definition it is
simple to derive expressions for the number of bonds, end-points, and
branch-points as derivatives of the logarithm of the grandpartition
function
\begin{align}
  \label{eq:Nb}
  N_b &= K \frac{\partial}{\partial K} \ln \Xi(K,f_1,f_3), \\
  \label{eq:N1}
  N_1 &= f_1 \frac{\partial}{\partial f_1} \ln \Xi(K,f_1,f_3), \\
  \label{eq:N3}
  N_3 &= f_3 \frac{\partial}{\partial f_3} \ln \Xi(K,f_1,f_3).
\end{align}
Following {the idea of de'Gennes \cite{deGennes:72a}, and using the methods of} Lubensky and Isaacson \cite{Lubensky:79a}, we equate the grand partition function for the branched polymers system with the $n\to0$ limit of the
partition function of an $\mathcal{O}(n)$ model of a magnet
\begin{equation}\label{eq:polymer_magnet_equiv}
\Xi(K,f_1,f_3; V) \approx 
\mathcal{Z}(K,f_1,f_3; V).
\end{equation}
The partition for the $\mathcal{O}(n)$ model of a magnet can be written as a function integral over a continuous field $\psi(x)$
\begin{equation}
\mathcal{Z}(K,f_1,f_3; V) = \int \!\! \mathcal{D} \psi \;
\exp \big( -\beta \mathcal{H}(\psi,K,f_1,f_3; V)\big).
\end{equation}
where $\mathcal{H}$ is the effective Hamiltonian
\begin{multline}\label{eq:effHam}
  \mathcal{H}(\psi,K,f_1,f_3;V) = \\
  \sum_{x}\bigg( \frac{1}{2} \vec{\psi}(\vec{x}) \sum\limits_{x'} 
  \delta_{\langle x, x' \rangle}^{-1}\cdot \vec{\psi}(\vec{x}')\\
  -\frac{K}{2} |\vec{\psi}(\vec{x})|^2 
  +\frac{K^2}{8}|\vec{\psi}(\vec{x})|^4 \\
  - f_1\sqrt{K} \psi_{1}(\vec{x}) -
  \frac{f_3 K^{\frac{3}{2}}}{6} \psi_{1}^3(\vec{x})\bigg).
\end{multline}
We emphasize here that the parameters $K$, $f_1$ and $f_3$ take on different meanings for the $\mathcal{O}(n)$ model of the magnet (see Appendix A). For example, the $K$ parameter represents the coupling constant between the nearest neighbors spins in the $\mathcal{O}(n)$ model of a magnet and the $\vec{\psi}$ field is the average magnetization in a small region for the magnet.
However, the $\vec{\psi}$ field in Eq.~\eqref{eq:effHam} is proportional to the monomer density for the branched polymer.
The first term in Eq.~\eqref{eq:effHam} is an entropic term that smoothes out the $\vec{\psi}(x)$ field. The $\delta_{\langle x, x' \rangle}^{-1}$ in the first term is the inverse of the nearest neighbor operator $\delta_{\langle x, x' \rangle}$, defined as 1 if $x$ and $x'$ are neighboring sites on a lattice and $0$ otherwise.  The second term in Eq.~\eqref{eq:effHam} proportional to $|\vec{\psi}|^2$ is due to the nearest neighbor attraction,  which tends to increase the $\vec{\psi}(x)$ field. The third term proportional to $|\vec{\psi}|^4$ is repulsive representing the self-avoiding nature of the polymer, and leads to a decrease in the $\vec{\psi}(x)$ field. The fourth and fifth terms proportional to $\psi_1(x)$ and $\psi^3_1(x)$ are both attractive, and show that both branch-points and endpoints serve to increase the $\psi(x)$ field. It is the balance of these attractive and repulsive terms that creates a well-defined finite $\psi(x)$ field in equilibrium.
For completeness, as well as pedagogical reasons, we
show the derivation of the equivalence between the grand canonical partition 
function, Eq.~\ref{eq:grand_partition}, for the polymer system and the partition function for the $n\to 0$
limit of the $\mathcal{O}(n)$ model and all relevant approximations for finding
the effective Hamiltonian, Eq.~\ref{eq:effHam}, in Appendix A.

We now make a mean field approximation and assume that the value of
the field $\psi(x)$ is uniform and is well approximated by its
average value $\psi_0$. Thus the sum over the nearest neighbors
simply becomes
\begin{equation}\label{eq:delta}
  \sum_{x}^V \delta_{\langle x x'\rangle} \psi(x') = z \psi_0,
\end{equation} 
with $z$ the number of nearest neighbors or coordination number. The
inverse of the nearest neighbor delta function is then simply the
reciprocal of the coordination number $\sum_{x}^V
\delta^{-1}_{\langle x x' \rangle} \psi(\vec{x}) =
\frac{1}{z}\psi_0$. So in the mean field theory, the grand canonical partition
function is the exponential of the effective Hamiltonian evaluated at
its minimum
\begin{equation}\label{eq:approx}
  \Xi(K,f_1,f_3;V) \approx e^{-\mathcal{H}(\psi_0,K,f_1,f_3;V)},
\end{equation}
with $\psi_0$ found by minimizing Eq.~\eqref{eq:effHam}, $\delta
\mathcal{H}/\delta \psi|_{\psi_0} = 0$,
\begin{equation}\label{eq:EOM1}
  \bigg(\frac{1}{z} - K\bigg)\psi_0 - f_1 \sqrt{K} - \frac{f_3
    K^{\frac{3}{2}}}{2} \psi_0^2 + \frac{K^2}{2}\psi_0^3 =0.
\end{equation}
From stastical mechanics we can identify the grand potential, $\beta \Phi_G(K,f_1,f_3;V)=-ln[\Xi(K,f_1,f_3;V)]$ using Eqs.~\eqref{eq:effHam}, \eqref{eq:delta} and \eqref{eq:approx} , 
\begin{multline}\label{eq:phi1}
  \beta \Phi_G(K,f_1,f_3;V) = \\ 
  \frac{V}{a^3}
  \Bigg[ \frac{1}{2}\bigg(\frac{1}{z}-K\bigg)\psi_0^2 
  - f_1 \sqrt{K}\psi_0 \\
  - \frac{f_3 K^{\frac{3}{2}}}{6} \psi_0^3 
  + \frac{K^2}{8}\psi_0^4 \Bigg],
\end{multline}
with $a$ the lattice spacing. Inserting Eq.~\eqref{eq:phi1} in
Eq.~\eqref{eq:Nb} and using Eq.~\eqref{eq:EOM1} we find the average
monomer density
\begin{equation}
  c_b = N_b/V = \frac{\psi^2_0}{2 z a^3}.
\end{equation}
At this point it is convenient to define a new field $\phi(x)$ such
that in the mean field approximation the average value $\phi_0^2 =
\psi_0^2/2 z a^3 $ is the monomer density. The grand potential written
in terms of the new field is
\begin{multline}\label{eq:PhiG1}
  \frac{\beta \Phi_G(K,F_1,F_3;V)}{V} = \\
  (1-zK)\phi_0^2  -f_1\sqrt{\frac{2zK}{a^3}} \phi_0 \\- \frac{1}{6}
  f_3(2 z a K)^{\frac{3}{2}} \phi_0^3 + \frac{1}{2} K^2 z^2 a^3
  \phi_0^4.
\end{multline}
By comparing Eq.~\eqref{eq:PhiG1} with the expression for the grand
potential for a linear polymer in good solvent \cite{Ji:88a}, we can
identify $r=(1-zK)$ as the chemical potential of monomers (such that
$\partial \beta \Phi_G / \partial r = c_b$) and $v=K^2 z^2 a^3$ as
the excluded volume. It is also convenient to absorb $\sqrt{ 2 z K}$ and $(2 z K)^{\frac{3}{2}}$ constants in the end and branch point fugacities, respectively such
that the grand potential can be written in a much simpler form
\begin{multline}\label{eq:PhiG2}
  \frac{\beta\Phi_G(K,f_1,f_3;V)}{V} = \\
  r \phi_0^2 - \frac{f_1}{\sqrt{a^3}} \phi_0 - 
  \frac{f_3}{6}\sqrt{a^3} \phi^3_0+\frac{v}{2}\phi_0^4.
\end{multline}
The parameters $f_1$ and $f_3$ are to physical quantities.
Using Eqs.~\eqref{eq:N1} and \eqref{eq:N3}, we find the
end-point and branch-point densities
\begin{align}
  c_1 &=  \frac{f_1}{\sqrt{a^3}} \phi_0, \label{eq:N1_2}\\
  c_3 &= \frac{1}{6}f_3 \sqrt{a^3}\phi^3_0.\label{eq:N3_2}
\end{align}

\section{Results and Discussion: Branched polymers adsorption onto different surfaces}\label{sec:RD}
We now apply the field theory presented in the previous section to a
semi-dilute system of annealed branched polymers and investigate their
adsorption to different surfaces.  More specifically, we consider
a solution of branched polymers with a monomer density $c_b$, where
the polymers all have a fixed length $L$, and a tunable average
branching number $N_b$.

The adsorption mean field energy of the branched polymer, $F-F_0$  can then be written as 
\begin{multline}
\label{energy1}
 F - F_0 = -\gamma a^3 \int d S \phi^2 + 
 \\ \int d V \bigg{(} \frac{ a^2 }{ 6 \beta} (\nabla \phi)^2+\frac{1}{2} \frac{\nu}{\beta} (\phi^4-c_b^2) \\
 - \frac{1}{\beta \sqrt{a^3}} (f_1 (\phi-c_b^{1/2}) +f_3 \frac{a^3}{6} (\phi^3-c_b^{3/2}))\bigg{)} 
\end{multline}
The first term in Eq.~\eqref{energy1} is a surface integral that
gives the contact energy between the surface and the polymer. The
first term in the volume integral is associated with the entropic
cost of a non-uniform polymer distribution\cite{Doi:86}. The rest of
the terms in Eq.~\eqref{energy1} are related to the free energy of a
branched polymer in mean field approximation, see
Eq.~\eqref{eq:PhiG2}. 
 
Considering the constraint that the total number of monomers is fixed
  \begin{equation}
\label{totaln}
\int dV \; \phi^2 = constant =N,
\end{equation}
Eq.~\eqref{energy1} can be rewritten as
\begin{multline}
\label{energy2}
 F - F_0 = -\gamma a^3 \int d S \phi^2 + \\
 \int d V \bigg{(} \frac{ a^2 }{ 6 \beta} (\nabla \phi)^2+\frac{1}{2} \frac{\nu}{\beta} (\phi^4-c_b^2)\\
  - \frac{1}{\beta \sqrt{a^3}} (f_1 (\phi-c_b^{1/2}) +f_3 \frac{a^3}{6} (\phi^3-c_b^{3/2})) -\lambda(\phi^2-c_b) \bigg{)}
\end{multline}
where $\lambda$ is the Lagrange multiplier.
  
Minimizing Eq.~\eqref{energy2} with respect to the field $\phi(x)$
gives the following Euler-Lagrange equation
\begin{equation}
\label{eulerl}
\frac{a^2}{6} \nabla^2 \phi= -\lambda \phi + \nu \phi^3 - \frac{1}{2 \sqrt{a^3}} (f_1 + \frac{a^3} {2} f_3 \phi^2)
\end{equation}
subject to the boundary condition
\begin{equation}
\label{bc}
\left[\hat{n}.\nabla \phi + ({6 \beta} {a} \gamma) \phi\right]_s=0
\end{equation}
For simplicity, we rescale the Lagrange multiplier $E= \frac{6
\lambda} {a^2}$ and introduce a length that characterizes the strength of the
attraction between the surface and monomers as $\kappa^{-1}=
\frac{1}{6 \beta a \gamma}$. The other boundary condition is natural, far
from the surface the field should be uniform, $\nabla ^2 \phi \rightarrow 0$, and and take on the bulk value $\phi(x) \rightarrow \sqrt{c_b}$. Using Eq.~\ref{eulerl}, the Lagrange multiplier can be
written as
\begin{equation} \label{epsilon}
  E=\frac {6} {a^2} (\nu c_b - \frac{1}{2} \frac{c_1}{c_b} 
  - \frac{3}{2} \frac{c_3}{c_b})
\end{equation}
where $c_1=f_1 c_b^{1/2}/\sqrt{a^3}$ and $c_3=f_3 \sqrt{a^3} c_b^{3/2}/6$ are respectively the end-point and branch-point concentrations
far from the surface. For long polymers with no loops in which $N_1=N_3+2$,
the second and third terms in Eq.~\eqref{epsilon} correspond to the ratio of
the end-points and branch-points to monomers. To make all the quantities
dimensionless, we rescale the field with respect to the bulk value $\theta(x)
= \phi(x) / \sqrt{c_b}$ and the spatial coordinate with respect to
the Edwards correlation length\cite{Edwards:65}  $x=\tilde{x} \xi_E$, where $\xi_E = \frac{a}{\sqrt{3 c_b v}}$, the
equation of motion simply becomes
\begin{equation}\label{thetaeulerl}
{\tilde{\nabla}}^2 \theta= 2 (\theta^3 -\theta)+A_1(\theta -1)+ 
3 A_3(\theta-\theta^2)
\end{equation}
with $A_1=(c_1/c_b)/(\nu c_b)$, $A_3=(c_3/c_b)/(\nu c_b)$.

The $A_1$ and $A_3$  quantities measure the relative importance of the branching structure of the polymer to the steric effect or  excluded volume interaction between monomers in solution . The numerators $c_1/c_b$ and $c_3/c_b$ are the ratios of the concentrations of end-points and branch-points to the total number of monomers, respectively. For large polymers, these ratios can approach $1/2$ for a maximally branched polymer. The denominator $v c_b$ is a filling fraction of the polymer; dilute solutions will have small values of $v c_b$, and a dense polymer system with no solvent will have a value of one.
 With the new scaled coordinates, the boundary conditions become
\begin{align}\label{thetabc}
  \left[{\frac{\partial \theta}{\partial \tilde{n}}
  +\tilde{\kappa}\theta} \right]_s=0 \\ \label{thetabc2}
  \lim_{x\to\infty} \theta = 1
\end{align}
with $\tilde{\kappa}=\kappa\xi_E$. In terms of the the new mean-field
order parameter, $\theta$, we can rewrite
Eq. \ref{energy2}, the adsorption energy,  as
\begin{multline}
\label{thetaenergy}
F-F_0= \frac{a^2}{6\beta}\xi_Ec_b \bigg( -\tilde{\kappa} \int d \tilde{S} \theta^2 \\
+ \int d \tilde{V} \big{(} (\tilde{\nabla}\theta)^2  + (\theta^2-1)^2 \\ -A_1(\theta-1)-{A_3}(2\theta^3-3\theta+1)\big{)}  \bigg)
\end{multline}
In the following sections we employ Eq.~\eqref{thetaenergy} to calculate the free energy of a polymer next to a flat, curved, spherical and sinusoidal surfaces. We note that while the Lagrange multiplier $\lambda$ acts like a chemical potential in open surfaces for fixing the density of bulk polymers, in the closed systems, like inside a sphere, it's used to fix the number of monomers inside the shell.

\subsection{{\bf{Analytical Calculations}}}
\subsubsection{Flat wall}

Next to a flat wall,  the Euler-Lagrange equation, Eq.~\eqref{thetaeulerl}, subject to the boundary conditions given in Eqs.~\eqref{thetabc} and \eqref{thetabc2} can be solved perturbatively. We assume the attractive interaction between monomers and wall is smaller than the monomer-monomer repulsion ($\tilde{\kappa}\ll 1$).  The solution to Eq.~\eqref{thetaeulerl} to the second order in $\tilde{\kappa}$ can then be written as

\begin{equation}
\label{flatsol}
\theta \approx 1 +  \frac{\tilde {\kappa}}{A} (1 + \frac{\tilde {\kappa}}{A}) e^{-A \tilde{z}} 
\end{equation}
where $A=\sqrt{4+A_1-3A_3}$ with $A_1$ and $A_3$ proportional to the number of end ($N_1$) and branch ($N_3$) points as given below Eq.~\eqref{thetaeulerl}. For a single long polymer with no loops Eq.~\eqref{eq:noloop} yields $N_1=N_3+2$. So we can simply write $A_1-3 A_3=2 V (1-N_3)/(\nu N^2)$ implying that $A_1-3A_3<0$ and $|A_1-3A_3|<4$ should be satisfied for real solutions. For $A_1=A_3=0$, Eq.~\eqref{flatsol} converges to the profile of a linear polymer next to the flat wall \cite{Ji:88a}. As clearly shown in Eq.~\eqref{flatsol},  the density of branched polymers are larger than the linear ones next to a flat wall.

Inserting Eq.~\eqref{flatsol} into Eq.~\eqref{thetaenergy}, we find the change in surface tension, energy per unit area, to the second order in $\tilde{\kappa}$

\begin{equation}
\label{flattension}
\gamma-\gamma_0 \approx \frac{{a^2} c_b}{6 \beta \xi_E} \big( - \tilde {\kappa} - \frac{\tilde {\kappa^2}}{2} - \Gamma_{b} (\tilde {\kappa},A_1,A_3) \big)
\end{equation}
where $\Gamma_{b}$ is the difference in tension from a linear chain due to branching and is given by 

\begin{multline}
\label{gammab}
\Gamma_{b}(\tilde {\kappa},A_1,A_3)=(\frac{A_1+3A_3}{A^2})\tilde{\kappa} \\ + (\frac{3A^2-A^3-4+2A_1+12A_3}{2 A^3})  \tilde{\kappa}^2
\end{multline}
Since the quantity $\Gamma_{b}$ for all acceptable values of $A_1$ and $A_3$ is always positive, according to Eq.~\eqref{flattension} the surface tension due to adsorption of a linear polymer to a flat wall is always higher than that of a branched one. In the next section, we calculate the impact of curvature on the polymer density profile and the free energy of the system.

\subsubsection{Curved wall}

To investigate the effect of curvature analytically, we assume that the flat wall is slightly bent to form a large sphere. The curvature could be either toward or away from the polymer. The radius of curvature $b$ is considered to be large compared to the correlation length ($\tilde{b}=b/\xi_E \gg 1$).
We can obtain the perturbative solutions in spherical coordinates through Eq. \ref{thetaeulerl} assuming that $\theta = 1+ \delta$ ($\delta \ll 1$) at the weak adsorption limit ($\tilde{\kappa}\ll1$). According to the direction of wall curvature, polymer solution is considered either to be in the interior (in) or the exterior (out) of a sphere of radius $b$. The perturbative solutions are then
\begin{equation}
\label{thetao}
  \theta_{out}(\tilde{r})\approx 1+ \frac{\tilde{\kappa} \tilde{b}}{1+A \tilde{b}-\tilde{\kappa}\tilde{b}} \big{(} \tfrac{\tilde{b}}{\tilde{r}}\big{)} e^{-A(\tilde{r}-\tilde{b})} 
\end{equation}
for $(\tilde{b}<\tilde{r}<\infty)$ and 
\begin{multline}
  \label{thetai}
   \theta_{in}(\tilde{r})\approx \\ 1+ \tilde{\kappa} \tilde{b} \big{(} \tfrac{\tilde{b}}{\tilde{r}}\big{)} \frac{\sinh A \tilde{r}}{A \tilde{b} \cosh(A \tilde{b})-\sinh (A \tilde{b})(1+\tilde{\kappa}\tilde{b})}
\end{multline}
for $(0<\tilde{r}<\tilde{b})$.
Asumming $\tilde{b}\gg 1$, we can write the monomer concentration on the surface to the second order in $\tilde{\kappa}$ 
\begin{align}
\label{concentrationout}
  \theta_{out}\approx 1+ \frac{\tilde{\kappa}}{A} \big{(} 1- \frac{1}{A \tilde{b}} +\frac{\tilde{\kappa}}{A} \big{)} &   \\
 \label{concentrationin}
   \theta_{in}\approx 1+ \frac{\tilde{\kappa}}{A} \big{(} 1+ \frac{1}{A \tilde{b}} +\frac{\tilde{\kappa}}{A} \big{)} & 
\end{align}
Comparison of Eqs.~\eqref{concentrationout} and \eqref{concentrationin} with Eq.~\ref{flatsol} reveals that branched polymer concentration in the vicinity of a flat wall increases if the wall bends toward the polymer and decreases if the wall bends away from the polymer, consistent with the results obtained for linear polymers \cite{Ji:88a}.
In order to obtain the change in surface tension due to the wall curvature, we insert the concentration profiles given in Eqs.~\eqref{thetao} and \eqref{thetai} into Eq.~\eqref{thetaenergy} and keep terms up to the first order in $1/\tilde{b}$. Then we have

\begin{multline}
\label{curvedtensionout}
  (\gamma-\gamma_0)_{out} \approx \\ 
  \gamma-\gamma_0 + \frac{a^2 c_b}{6 \beta \xi_E} \Big {(} 
  \frac{\tilde{\kappa}^2}{4 \tilde{b}} +
  \Gamma_{g} (\tilde{\kappa},\tilde{b},A_1,A_3) \Big{)}
\end{multline}
and
\begin{multline}
  \label{curvedtensionin}
  (\gamma-\gamma_{0})_{in} \approx \\ 
  \gamma-\gamma_0 - \frac{a^2 c_b}{6 \beta \xi_E} \Big {(} 
  \frac{\tilde{\kappa}^2}{4 \tilde{b}} +
  \Gamma_{g} (\tilde{\kappa},\tilde{b},A_1,A_3) \Big{)}
\end{multline}
where $\gamma-\gamma_0$ is the surface tension for the polymer next to a flat wall based on Eq.~\eqref{flattension} and $\frac{\tilde{\kappa}^2}{4 \tilde{b}}$ is the difference in surface tension due solely to the geometry of the wall. The quantity 
\begin{equation}
\label{gammag}
\Gamma_g(\tilde{\kappa},\tilde{b},A_1,A_3) = \frac{8 A^2 - A^4 -16 +4 A_1 + 36 A_3}{4 A^4 \tilde{b}} \tilde{\kappa}^2
\end{equation}
is due to the the coupling between the geometry and branched structure of the polymers.
If we set $A_1=0$ and $A_3=0$ then $A=2$ and Eqs.~\eqref{curvedtensionout} and \eqref{curvedtensionin} reveal the impact of wall curvature on the surface tension for the case of linear polymers. As expected, the surface tension decreases if the wall bends toward the polymer and it increases if the wall bends away from the polymer.

 Quite interestingly, the $\Gamma_g$ expression shows the importance of coupling between wall curvature and polymer branching on the surface tension. A glance through Table 1.~shows that the sum or difference $\Gamma_{b} \pm \Gamma_g$ corresponds to the change in the surface tension due to the coupling between branching and wall curvature. Since this term is positive for both the convex and concave surfaces, we see that branching always decreases the surface tension. However, the coupling between branching and curvature further lowers the surface tension when the wall curves toward the polymers compared to when the wall curves away from the polymer. 

\begin{table} 
{\bf
\begin{tabular}{|l|c|c|}
 \hline
 & Concentration  & Surface Tension \\ \hline\noalign{\smallskip}
 Flat surface & 
 $ \theta_{\text{flat}} = 1+ 
 \frac{\widetilde{\kappa}}{A}\big(1+\frac{\widetilde{\kappa}}{A}) $ & 
 $ \Delta \gamma_{\text{flat}} \approx 
 \frac{{a^2} c_b}{6 \beta \xi_E} 
 \big( - \tilde{\kappa} - \frac{ \tilde{\kappa^2}}{2} 
 - \Gamma_{b} \big)$ \\ 
 \noalign{\smallskip}\hline\noalign{\smallskip}
 Convex surface & 
 $ \theta_{\text{out}} = \theta_{\text{flat}} -
 \frac{\widetilde{\kappa}}{A^2 \widetilde{b}} $ & 
 $ \Delta \gamma_{\text{out}} \approx 
 \Delta \gamma_{\text{flat}} + \frac{{a^2} c_b}{6 \beta \xi_E} 
 \big( \frac{\widetilde{\kappa}^2}{4\widetilde{b}} + \Gamma_{g}\big) $ \\
 \noalign{\smallskip}\hline\noalign{\smallskip}
 Concave surface & 
 $ \theta_{\text{in}} = \theta_{\text{flat}} +
 \frac{\widetilde{\kappa}}{A^2 \widetilde{b}} $ & 
 $ \Delta \gamma_{\text{in}} \approx 
 \Delta \gamma_{\text{flat}} -  \frac{{a^2} c_b}{6 \beta \xi_E} 
 \big( \frac{\widetilde{\kappa}^2}{4\widetilde{b}} + \Gamma_{g}\big) $ \\ 
 \noalign{\smallskip}\hline
\end{tabular}
\caption{
  \label{tab:summary} Summary of the analytic results. The table shows the impact of the wall geometry and the topology of the polymer (embedded in $\Gamma_{g}$ and $\Gamma_{b}$, see the text) for two physical quantities of $\theta$ and $\Delta \gamma$. The quantity $\theta$ is related to the concentration of the polymer at the surface by $c_b \theta^2$ with $c_b$ monomer density. $\Delta \gamma$ is the change in the surface tension due to the presence of the wall.
  }
}
\end{table}


\subsection{{\bf{Numerical Calculations}}}
\label{sec:num}

The above analytical calculations were related to the surfaces with large radius of curvature. To study polymer adsorption on the surfaces with higher curvature, we need to numerically solve Eq.~\eqref{thetaeulerl} for both the interior and exterior of smaller spheres. Since this is mainly a phenomenological model, we follow similar parameters to previous numerical works and values that are typical for virus capsids and RNA sizes \cite{Garmann,Comas,Ji:88a,Vanderschoot2009,bruinsma,Vanderschoot2005,Anze}.

\subsubsection{Outside the sphere}

\begin{figure}
\begin{center}
\includegraphics[width=7.0cm,height=5cm]{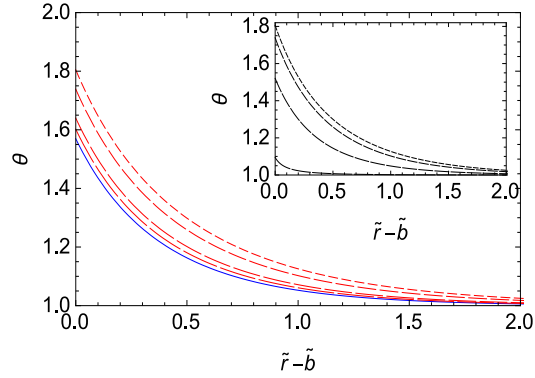}
\caption{Profiles of the scaled polymer density amplitude $\theta$ vs scaled distance from the surface of the sphere $\tilde{r}-\tilde{b}$ for several relative branching density $A_3$. The solid line gives the profile for a linear polymer with no branching, and dashed lines give the profiles for branched polymers with branching density $A_3=0.2, 0.4, 0.8, 1.0$. Other parameters used are $\tilde{\kappa}=1$, $\tilde{b}=5.0$ and $\nu=0.5$. The inset of Fig.~\ref{thetaout} is the concentration profile for several values of the radius of the sphere ($\tilde{b}=0.1, 1, 5, 100$) for the branching density $A_3=0.8$ with $\tilde{\kappa}=1$ and $\nu=0.5$. As the radius of the sphere increases, monomer concentration on the surface increases.}
\label{thetaout}
\end{center}
\end{figure}
\begin{figure}
\begin{center}
\includegraphics[width=7.0cm,height=5cm]{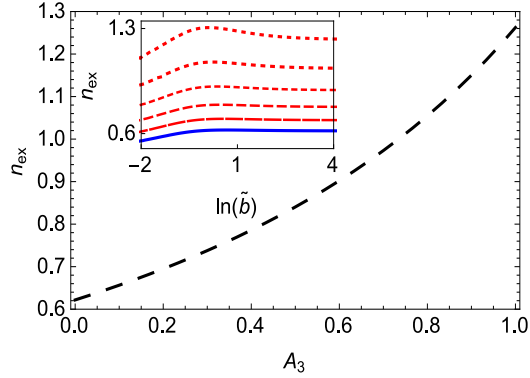}
\caption{Surface excess (in units of $\xi_E c_b$) vs branching density for $\tilde{\kappa}=1.0$, $\tilde{b}=5.0$ and $\nu=0.5$. Inset shows the excess on different size of spheres for $\tilde{\kappa}=1.0$ and $\nu=0.5$. The solid line represents the linear chain with $A_3=0$, and the dashed lines give the surface excess for branched polymers with branching density $A_3=0.2, 0.4, 0.6, 0.8, 1.0$ as dashing gets smaller. }
\label{Nex}
\end{center}
\end{figure}

In this section, we consider smaller spheres and obtain the polymer
concentration profile outside the sphere vs the scaled distance from
the surface of the sphere, $\tilde{r}-\tilde{b}$. We obtain numerical
results by solving the nonlinear differential equation Eq.~\eqref{thetaeulerl} subject to the boundary conditions given in Eqs.~\eqref{thetabc} and \eqref{thetabc2}. The results are presented in
Fig.~\ref{thetaout}, which illustrates that both the monomer
concentration at the wall and the thickness of the adsorption layer
increase as the branching density increases.  The surface excess adsorbed onto the sphere can also be calculated using the
concentration profile $\theta$,

\begin{equation}
\label{surfaceexcess}
n_{ex} = \xi_E c_b \int _{\tilde{b}}^{\infty} d \tilde{r} (\theta^2-1)(\frac{\tilde{r}}{\tilde{b}})^2
\end{equation}
Figure~\ref{Nex} shows the surface excess as a function of the branching density. As the branching density increases, the surface excess also increases. For $A_3=0.2, 0.4, 0.6, 0.8, 1.0$, the ratio of surface excess of a branched to a linear polymer is $n_{ex}^{branched}/n_{ex}^{linear}=$1.118, 1.264, 1.451, 1.696, 2.030, respectively.  Inset of Fig.~\ref{Nex} is a plot of surface excess vs the radius of the sphere for different branching densities $A_3=$0.2, 0.4, 0.6, 0.8, 1.0. According to the inset of Fig.~\ref{Nex}, for a given branching density, there is a critical radius $r^*$ for which the surface excess has a maximum. We find that the position of the critical radius decreases linearly as branching density goes up, {\it i.e., } $r^* \propto -A_3$.

\begin{figure}
\begin{center}
\includegraphics[width=7.0cm,height=5cm]{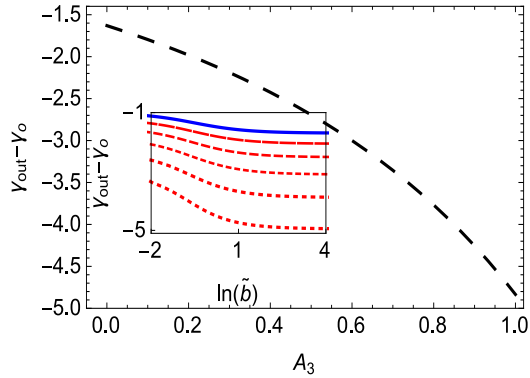}
\caption{Surface tension (in units of $\frac{a^2 c_b}{6 \beta \xi_E}$ ) vs branching density for $\tilde{\kappa}=1.0$, $\tilde{b}=5.0$ and $\nu=0.5$. Inset is a plot of surface tension as a function of radius of the sphere for $\tilde{\kappa}=1.0$ and $\nu=0.5$. The solid line represents the linear chain with $A_3=0$, and the dashed lines give the surface tension for branched polymers with branching density $A_3=0.2, 0.4, 0.6, 0.8, 1.0$ as dashing gets smaller. }
\label{tension}
\end{center}
\end{figure}

Using the numerical solution for $\theta$ and Eq.~\eqref{thetaenergy}, the surface tension can be written as
\begin{multline}
\label{tensionsphericalout}
\gamma_{out}-\gamma_{0}=\frac{a^2 c_b}{6 \beta \xi_E} \bigg( - \tilde{\kappa} \theta^2 (\tilde{b}) \\
 + \int_{\tilde{b}}^{\infty} dr \big( (\theta^2-1)^2-A_1(\theta-1)
 \\ -{A_3}(2 \theta^3-3 \theta+1) +(\tilde{\nabla}\theta)^2   \big) (\frac{\tilde{r}}{\tilde{b}})^2 \bigg).
\end{multline}
Figure~\ref{tension} is a plot of the surface tension as a function of the branching density, which shows that as the branching density increases, the surface tension decreases. The inset shows the tension vs the sphere radius for different branching densities $A_3=$0.2, 0.4, 0.6, 0.8, 1.0. As the radius of sphere increases,  the tension decreases consistent with the perturbative results.

\subsubsection{Inside a sphere}

\begin{figure}
\begin{center}
\includegraphics[width=7.0cm,height=5cm]{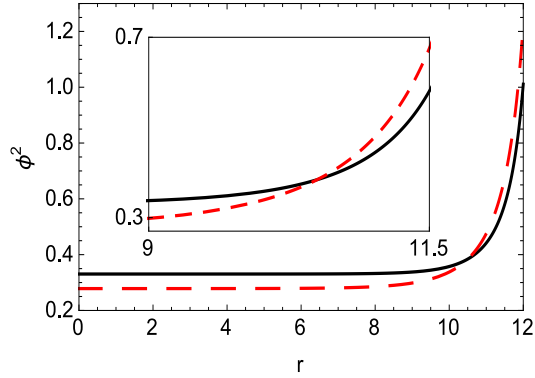}
\caption{Concentration profile ($ \phi(x)=\theta(x) \sqrt{c_b}$) inside the sphere for branching density $A_3=0.05$ (solid line) and  $A_3=1.04$ (dashed lines) for $b=12$, $\tilde{\kappa}=1.0$, $\nu=0.5$ and $N=3000$ where the lengths are in units of a, the Kuhn length. Inset shows the details of the plot next to the surface.}
\label{densityin}
\end{center}
\end{figure} 

\begin{figure}
\begin{center}
\includegraphics[width=7.0cm,height=5cm]{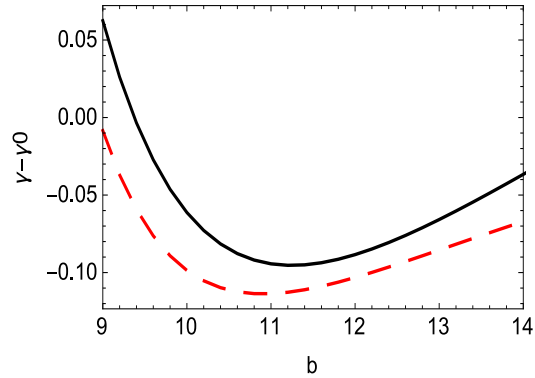}
\caption{ Surface tension for polymer adsorbed in the interior of the sphere at  $A_3=0.05$ (solid line) and  $A_3=1.04$ (dashed lines) for $\tilde{\kappa}=1.0$, $\nu=0.5$ and $N=3000$ where the lengths are in units of a, the Kuhn length and the energy is in units of $k_BT$.}
\label{tensionin}
\end{center}
\end{figure} 
In this section, we obtain the polymer
concentration profile inside a sphere ($r<b$) by solving the nonlinear differential equation Eq.~\eqref{thetaeulerl} subject to the boundary conditions given in Eq.~\eqref{thetabc}. In addition, because the polymer is now confined inside an impermeable sphere, the total number of monomers, N is fixed. In terms of normalized length scale and the order parameter $\theta$, we have
\begin{equation}
\label{constraint}
N = c_b \xi_E^3 \int_0^{\tilde{b}} 4 \pi \tilde{r}^2 \theta^2 d \tilde{r}.
\end{equation}
The surface tension in this case can be written as
\begin{multline}
\label{tensionsphericalin}
\gamma_{in}-\gamma_{0}=\frac{a^2 c_b}{6 \beta \xi_E} \bigg( - \tilde{\kappa} \theta^2 (\tilde{b}) \\
 + \int_{0}^{\tilde{b}} dr \big( (\theta^2-1)^2-A_1(\theta-1) \\ -{A_3}(2 \theta^3-3 \theta+1) +(\tilde{\nabla}\theta)^2   \big) (\frac{\tilde{r}}{\tilde{b}})^2 \bigg).
\end{multline}

The concentration profile as a function of r, the distance from the
center of the sphere is shown in Fig.~\ref{densityin} for a branched
polymer with branching density $A_3=0.05$ (solid lines) and
$A_3=1.04$ (dashed lines). As illustrated in the figure, due to the
attraction between the wall and the polymer, more monomers are
attracted to the surface of sphere as the branching density
increases.

Figure~\ref{tensionin} illustrates the surface tension as a function
of the radius of the sphere. Quite interestingly, we find that there
is an optimal size for the radius of sphere for a given fixed chain
length.  For the parameters $\tilde{\kappa}=1.0$, $\nu=0.5$ and
$N=3000$, the optimal radius is $b=11.25$ with $A_3=0.05$  and  is
$b=10.90$ for $A_3=1.04$. When the polymer is more branched, the
optimal radius becomes smaller. This is mainly due to lower cost for
the excluded volume interaction.

\subsubsection{Sinusoidal grating}

\begin{figure}
\begin{center}
\includegraphics[width=7.0cm,height=5cm]{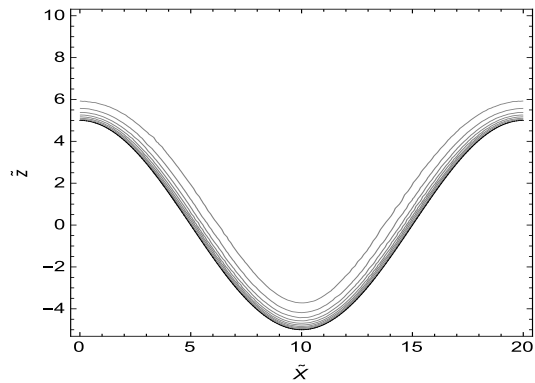}
\caption{Contour plot of the concentration profile for the polymer with branching density $A_3=0.8$ for $\tilde{\kappa}=1$, $\tilde{\lambda}=20$ and $\tilde{z_0}=5$.  The solid thick black line is the position of the surface. The upper contour corresponds to a local concentration of $\theta^2=1.25$ that increases to $\theta^2=3.6$ towards the bottom of the valley. The $\tilde{x}$ and $\tilde{z}$ coordinates are distances along and perpendicular to the corrugations respectively. }
\label{lambda20}
\end{center}
\end{figure}
\begin{figure}
\begin{center}
\includegraphics[width=7.0cm,height=5cm]{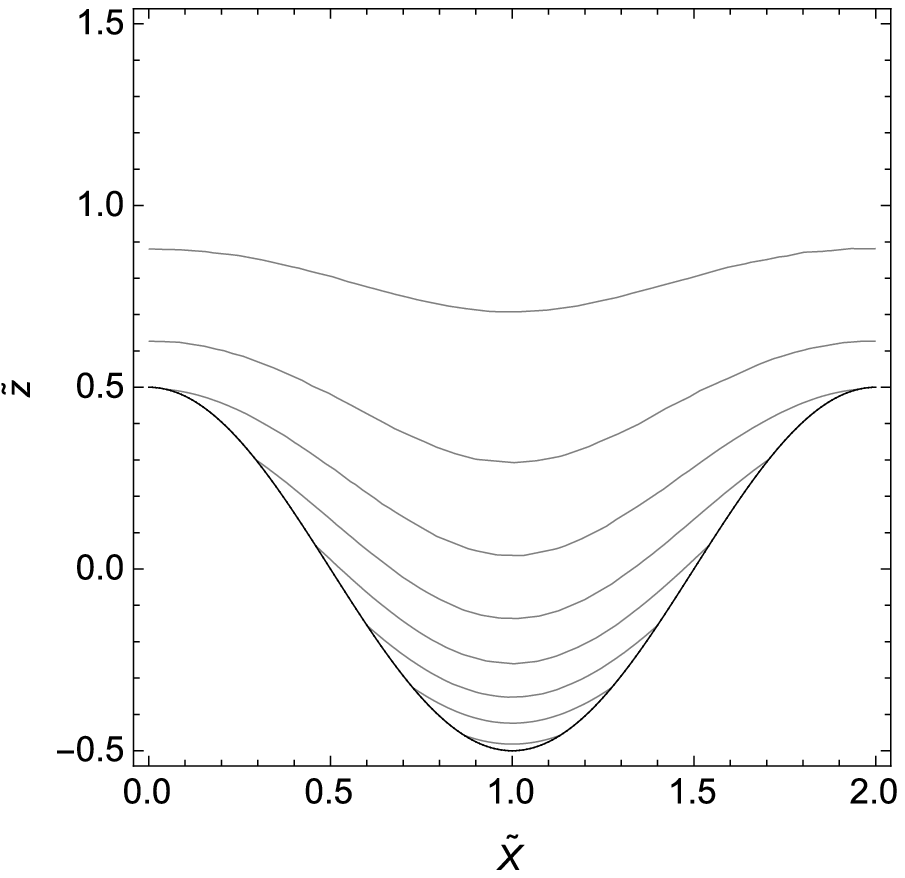}
\caption{Contour plot of the concentration profile for the polymer with branching density $A_3=0.8$ for $\tilde{\kappa}=1$, $\tilde{\lambda}=2$ and $\tilde{z_0}=0.5$.  The solid thick black line is the position of the surface. The upper curve corresponds to a local concentration of $\theta^2=1.5$ that increases to $\theta^2=5$ towards the bottom of the valley. The $\tilde{x}$ and $\tilde{z}$ coordinates are distances along and perpendicular to the corrugations respectively.} 
\label{lambda2}
\end{center}
\end{figure}
\begin{figure}
\begin{center}
\includegraphics[width=7.0cm,height=5cm]{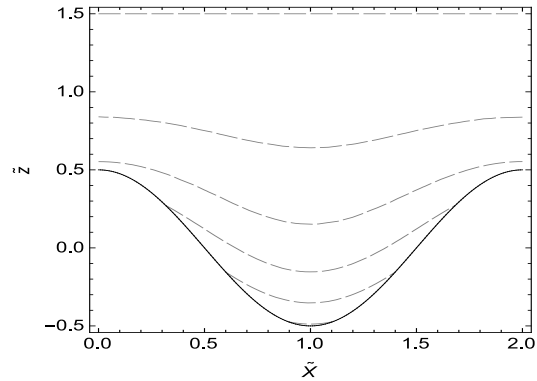}
\caption{Contour plot of the ratio of branch to monomer concentrations with branching density $A_3=0.8$ for $\tilde{\kappa}=1$, $\tilde{\lambda}=2$ and $\tilde{z_0}=0.5$. The solid thick black line is the position of the surface. The upper curve corresponds relative concentration of branch points of $c_3/c=0.4$ and it increases to $c_3/c=0.9$ towards the bottom of the valley.  The $\tilde{x}$ and $\tilde{z}$ coordinates are distances along and perpendicular to the corrugations respectively. }
\label{branchratio}
\end{center}
\end{figure}

In this section, we consider a polymer solution next to a sinusoidal
surface, $z = z_0 cos ((2\pi/\lambda)x)$. Here $z_0$ is the amplitude
and $\lambda$ is the wavelength of the surface.  As mentioned in the
introduction, this should give some insight into the behavior of
branched polymer next to a rough random surface because the
qualitative features of the results are the same. To obtain the
concentration profile, $\theta$ we solve Eq.~\eqref{thetaeulerl}
subject to the boundary condition given in Eq.~\eqref{thetabc} using
a finite element method in 2D. The numerical results
are shown in Figs.~\ref{lambda20}, \ref{lambda2} and
\ref{branchratio} as contour plots of the polymer density next to the sinusoidal adsorbing surface.  In all cases we keep the strength of the
attractive interaction between the surface and monomers the same.

Figure~\ref{lambda20} shows that the profile of a branched polymer
next to a sinusoidal surface is similar to that of a flat wall if the
wavelength of surface fluctuations is large, for example
$\tilde{\lambda}=\lambda/\xi_E=20$. The figure also illustrates that
the concentration of genome is higher in the valley compared to the
peaks. This is consistent with the perturbative results presented in
previous section, in that if the wall curves away from the
genome, the monomer concentration decreases, otherwise, it increases.
The non-uniformity in the concentration profile at the wall becomes
more apparent as we decrease $\tilde{\lambda}$; {\it i.e.,} the
genome concentration becomes much higher at the valley compared to
the peak, see Fig.~\ref{lambda2}. In the figure, the amplitude of
surface fluctuations, $\tilde z_0=z/\xi_E=0.5$, is chosen to be relatively
small to emphasize on the difference between the genome profile next
to the flat and sinusoidal walls. Note that the amplitude $\tilde
z_0= 0.5$ in Fig.~\ref{lambda2} is 10 times smaller than that in
Fig.~\ref{lambda20}. Nevertheless, since
$\tilde{\lambda}=2$ is 10 times smaller in
Fig.~\ref{lambda2}, the impact of surface fluctuations are more
pronounced.

In addition, we find that not only the concentration profile
at the wall is not uniform, the distribution of branch-points is not
homogeneous either. Figure \ref{branchratio} illustrates the ratio of branch
density to the monomer density at a sinusoidal surface. The figure shows that
the branching concentration is higher at the valley with respect to peaks
consistent with our perturbative results in section 2, where we found that the
branching density increases if the surface is curved toward the polymer and
decreases if the surface is curved away.

\section{Conclusion}\label{sec:con}

In this paper, we use the field theory methods based on the $n\to0$
limit of an $\mathcal{O}(n)$ model to describe the statistics of an
annealed branched polymer. We, in particular, carefully examine the behavior of branched polymers next to various adsorbing walls both analytically and numerically.

We show that for the annealed branched polymers, increasing the branching density will increase the concentration of polymer but decrease the surface tension next to flat, inward curving, and outward curving walls. In comparison to a flat adsorbing wall, we find that when the wall curves toward the polymer solution, the tension decreases but when it curves away from it, the tension increases. While these results are consistent with those found for linear polymers next to different type of walls, we interestingly found a correlation between the branching and curvature that causes a further lowering of surface tension when the wall curves towards the polymer, but decreases the amount of lowering of surface tension when the wall curves away from the polymer.

Our numerical solutions for the adsorption of polymer to the exterior of small spheres show that increasing the branching lead to an increase in the surface excess. For the polymers adsorbed in the interior of small spheres, we find a minimum in the surface tension as a function of radius of sphere.  This clearly demonstrates the interplay between monomer-wall attraction and monomer-monomer excluded volume interaction. Our findings also indicate that branching decreases the optimal radius of sphere, as more polymers can sit in the vicinity of the wall without a huge cost for monomer-monomer repulsion. This result has a considerable consequence for the encapsulation of RNA by virus shell proteins, and could suggest a non-specific mechanism for the preferential packaging of viral RNA to cellular RNA {\it in vivo} \cite{Erdemci:13a}.

Furthermore, we find similar effect for the adsorption of polymers onto sinusoidal surfaces. The concentration of polymers increases in the valleys compared to peaks and also branching density goes up in the valley section compared to the peaks. This effect becomes more pronounced as wave-length decreases. 

Understanding the mechanisms involved with the adsorption of annealed branched polymers onto different surfaces will play a critical role in biomedical technologies. In particular, the paper was inspired by the idea of
using functionalized gold nano-particles to bind RNA for gene
delivery\cite{Delong:10a}, which has industrial applications for
biosensors and microfluidic devices, and even possible medical
application for gold nano-particles encapsulated by virus coats as
potential tools for gene therapy. \cite{Sun:07a, Delong:10a,Johnson:12a,Rothberg:05a,ChenDragnea2006,SiberZandi2010}.

\section*{Acknowledgment}
This work was supported by the
National Science Foundation through Grant No. DMR-13-10687.

\appendix
\setcounter{section}{1}
\section*{Appendix A: $O(n)$ model of a magnet}

In this appendix, we derive the equivalence between the grand canonical partition 
function, Eq.~\ref{eq:grand_partition}, for the polymer system and the partition function for the $n\to 0$
limit of an $\mathcal{O}(n)$ model. Note that the $\mathcal{O}(n)$ model corresponds to a magnet whose magnetic
dipole of its atoms has $n$ components . The Ising model commonly studied in most statical mechanics
courses is the O(n) model for n=1. The n=0 limit is an interesting case
as it reproduces the statistics of a self-avoiding linear
polymer.

The Hamiltonian of the $O(n)$ model of a magnet is
\begin{multline} \label{eq:hamil}
  H(\{\vec{S}\},K,f_1,f_3;V) = - K \sum_{\langle x, x' \rangle}^V
  \vec{S}_x \cdot \vec{S}_{x'} \\ - f_1 \sum_{x}^V J_1[\vec{S}_x] -
  f_3 \sum_{x}^V J_3[\vec{S}_x],
\end{multline}
where $S_x$ is an $n$ dimensional vector of fixed length at each
lattice point $x$ and $K$, $f_1$, and $f_3$ are the coupling constants.
The first sum in Eq.~\ref{eq:hamil} is over all pairs of nearest neighbors and
$J_1[\vec{S}_x]$ and $J_3[\vec{S}_x]$ are the source terms for end-points and branch-points respectively. We have used some prescience
in giving the coupling constants $K$, $f_1$, and $f_3$ the same
symbol as the fugacities in Eq.~\eqref{eq:grand_partition}. The
partition function for the $\mathcal{O}(n)$ model is then
\begin{equation}\label{eq:Z1}
  \mathcal{Z}_n(K,f_1,f_3;V) = \prod_{x}^V \tr_{\vec{S}_x}
  e^{-H(\{\vec{S}\},K,f_1,f_3;V)},
\end{equation}
where $\tr_{\vec{S}}$  defined as 

\begin{equation}
\label{trace_def}
\tr_{\vec{S}} \; e^{\vec{k}.\vec{S}}= \frac    {\prod \limits_{i} (\int \limits_{-\infty}^{\infty} dS_i)   \delta(\sum \limits_i S_i^2-n)) e^{\vec{k}.\vec{S}} }     {\prod \limits_{i} (\int \limits_{-\infty}^{\infty} dS_i)   \delta(\sum \limits_i S_i^2-n)) }
\end{equation} 

and the size of the spin is subject to the condition $|\vec{S}|=\sqrt{n}$ or $\sum \limits_i S_i^2=n$.

Using the power series
definition of the exponential, Eq.~\eqref{eq:Z1} can be written as
\begin{multline}\label{eq:Z2}
  \mathcal{Z}_n(K,f_1,f_3;V) = \\ 
  \sum_{N_b=0}^{\infty} \frac{K^{N_b}}{N_b!} 
  \sum_{N_3=0}^\infty \frac{f_3^{N_3}}{N_3!}
  \sum_{N_1=0}^\infty \frac{f_1^{N_1}}{N_1!} 
  I_n(N_b,N_3,N_1;V),
\end{multline}
with $I_n$ defined as
\begin{multline}\label{eq:Idef}
  I_n(N_b,N_3,N_1;V) = \\ \prod_{x}^V \tr_{\vec{S}_x} 
  \Bigg[\!\! \bigg(\sum_{\langle x,x'\rangle}^V  
  \vec{S}_{x} \cdot \vec{S}_{x'}\bigg)^{\!N_b} \\
  \bigg(\sum_x^V J_3[\vec{S}_x]\bigg)^{\!N_3} \!\!
  \bigg(\sum_x^V J_1[\vec{S}_x]\bigg)^{\!N_1}\!\! \Bigg] = \\ \prod_{x}^V \tr_{\vec{S}_x} 
  N_b! N_3! N_1!\sum_\alpha \mathcal{C}_\alpha[\{\vec{S}\}],
\end{multline}
The comparison of the grand canonical partition functions defined in
Eq.~\eqref{eq:grand_partition} with the partition function for the
$\mathcal{O}(n)$ model in Eq.~\eqref{eq:Z2} reveals the similarity
between the two models.  It is now obvious why the coupling constants
in the $\mathcal{O}(n)$ model were chosen to be labeled the same as
the fugacities in the polymer system on a lattice with the excluded
volume interaction. 
To show full equivalence between the grand canonical ensemble of self-avoiding branched polymers $\Xi$ in Eq.~\eqref{eq:grand_partition} and the $n\to0$ limit of partition function for the $O(n)$ model of a magnet (Eq.~\eqref{eq:Z1}, we only need to show that in the $n\to 0$ limit the $I_n$ expression gives the multiplicity $\Omega$, or counts the number of ways of arranging a self-avoiding branched polymer on the lattice. 

Each $\mathcal{C}_\alpha[\{\vec{S}\}]$ in Eq.~\eqref{eq:Idef} can be
represented graphically.  For the lattice, a product of neighboring
$\vec{S}$-vectors that lie on points $x$ and $x'$ can be represented
by a line drawn between the points. The $J_1[\vec{S}_x]$ and
$J_3[\vec{S}_x']$ source terms at point $x$ and $x'$ can be
represented by circles and triangles placed on their respective
points.
\begin{figure}
\includegraphics[width=0.2\textwidth]{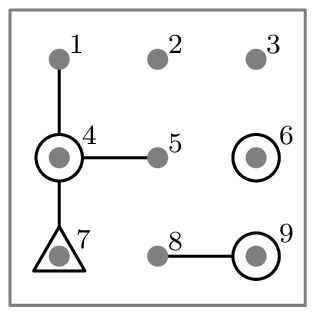}
\includegraphics[width=0.2\textwidth]{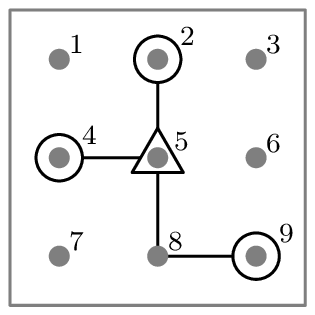}
  \caption{\label{fig:fig3} Two possible configurations of 4 bonds, 3
    end-points, and 1 branch-point on a $3\times3$ two-dimensional
    square lattice. The left diagram is not a valid configuration
    while the right diagram is a valid configuration and is counted.}
\end{figure}

As an example we present two possible configurations, one that represents a valid branched polymer configuration and one that does not, with $N_b=4$,
$N_1=3$, and $N_3=1$ on a $3 \times 3$ square lattice in
Fig.~\ref{fig:fig3}.  The
corresponding $\mathcal{C}_\alpha[\{\vec{S}\}]$ terms for both graphs in
Fig.~\ref{fig:fig3} contain 4 dot products of neighboring
$\vec{S}$-vectors (one for each bond), 3 $J_1[\vec{S}]$ terms (3
circles), and 1 $J_3[\vec{S}]$ term (1 triangle). The
$\mathcal{C}_\alpha[\{\vec{S}\}]$ term for the right diagram  in
Fig.~\ref{fig:fig3} is
\begin{multline}\label{eq:fig3_Cb}
  \mathcal{C}_b[\vec{S}] =
  \big(\vec{S}_2 \cdot \vec{S}_5\big)
  \big(\vec{S}_4 \cdot \vec{S}_5\big)
  \big(\vec{S}_5 \cdot \vec{S}_8\big)\\
   \big(\vec{S}_8 \cdot \vec{S}_9\big)
  J_1[\vec{S}_2]J_1[\vec{S}_4]
  J_1[\vec{S}_9]J_3[\vec{S}_5].
\end{multline}
where the indices indicate which lattice site is associated with each
term.

We will show below that in the $n\to0$ limit the trace of a
single configuration gives one if the graph corresponds to a physically valid branched polymer
configuration and zero otherwise, i.e.,
\begin{equation}\label{eq:C1}
  \lim_{n\to 0}\prod_x^V \tr_{\vec{S}_x} \mathcal{C}_\alpha[\{\vec{S}\}]
  =\begin{cases} 1 &\text{ for $\alpha$ valid,}\\
  0 &\text{ for $\alpha$ invalid.}\end{cases}
\end{equation}
This indicates, in the $n\to0$ limit, the $I_n$ term given in Eq.~\eqref{eq:Z2} counts
the number of valid physical configurations, which is how multiplicity
$\Omega$ is defined in Eq.~\eqref{eq:grand_partition}. The
condition in Eq.~\eqref{eq:C1} establishes the equivalence between the
partition function for the $\mathcal{O}(n)$ model in Eq.~\eqref{eq:Z2}
with the grand canonical partition function for a flexible branched
polymer in Eq.~\eqref{eq:grand_partition}, {\it i.e.}
\begin{equation}
  \lim_{n\to0} \mathcal{Z}_n(K,f_1,f_3;V) =
  \Xi (K,f_1,f_3;V).
\end{equation}

\subsection*{Evaluation of the trace}\label{sec:integral}
The trace of the configurations $\mathcal{C}_\alpha[\{\vec{S}\}]$ in
Eq.~\eqref{eq:C1} takes the form of products of $\vec{S}$-vectors on
each lattice site $x$. These products can be evaluated by using
the following generating function
\begin{equation}\label{eq:gen1}
  \tr_{\vec{S}} S_{i_1} \cdots S_{i_p} = \frac{\partial}{\partial
    k_{i_1}} \cdots \frac{\partial}{\partial k_{i_p}} \tr_{\vec{S}}
  e^{\vec{k} \cdot \vec{S}}
  \!\!\!\!\!\operatorname*{\Big|}_{\vec{k}\to 0}\!\!\!\!\!.
\end{equation}
The trace of the generating function $\tr_\vec{S}
e^{\vec{k}\cdot\vec{S}}$ can be evaluated in closed form 
\begin{equation}\label{eq:gen4}
  \tr_\vec{S} e^{\vec{k}\cdot\vec{S}} =
  \sum_{p=0}^\infty 
  \frac{\Gamma(\frac{n}{2})}{\Gamma(\frac{n}{2}+p)}\frac{1}{p!}
  \bigg(\frac{n |\vec{k}|^2}{4}\bigg)^p.
\end{equation}
In the limit of $n\to0$ it simplifies to the form
\begin{equation}\label{eq:gen3}
  \lim_{n\to0} \tr_\vec{S} e^{\vec{k}\cdot\vec{S}} =
  1 + \frac{1}{2} |\vec{k}|^2.
\end{equation}
A detailed step by step derivation of Eq.~\eqref{eq:gen4} is presented in appendix A of Ref.~\cite{Zilman:02a}. Inserting Eq.~\eqref{eq:gen3} into
Eq.~\eqref{eq:gen1}, we find the product of the components of the
$\vec{S}$-vectors in the absence of source terms is
\begin{equation}\label{eq:Sint}
  \lim_{n\to0} \tr_{\vec{S}} S_{i_1} \cdots S_{i_p} =
  \begin{cases} 
    1 &\quad\text{if } p=0,\\
    \delta_{i_1 i_2} &\quad\text{if } p=2,\\
    0 &\quad\text{otherwise.}
  \end{cases}
\end{equation}
The expression in Eq.~\eqref{eq:Sint} evaluates to a non-zero value
only in the lattice sites with exactly 0 or 2 bonds terminating on
them. 

To describe the generating function for end- and branch-points, we
construct $J_1$ and $J_3$ functions, respectively, such that they
satisfy the following equations
\begin{subequations}\label{eq:Jints}
\begin{align}
  \lim_{n\to0} \tr_{\vec{S}} S_i J_1[\vec{S}] & = 
  \lim_{n\to0} \frac{1}{\sqrt{n}} \sum_j\delta_{ij},\label{eq:J1int}\\
  \lim_{n\to0} \tr_{\vec{S}} S_i S_j S_k J_3[\vec{S}] & = \label{eq:J3int}
  \lim_{n\to0} \sqrt{n} \sum_{l} \delta_{il}\delta_{jl}\delta_{kl}, 
\end{align}
\end{subequations}
while all other traces involving the sources such as $\tr_{\vec{S}}
S_i S_j J_1[\vec{S}]$, $\tr_{\vec{S}} S_i J_3[\vec{S}]$ and
$\tr_{\vec{S}} S_i S_j J_3[\vec{S}]$ are equal to zero in the $n\to0$
limit.  Using Eq.~\eqref{eq:gen4} it is straightforward to derive the
following expressions for $J_1$ and $J_3$,
\begin{subequations}\label{eq:Jdefs}
\begin{align}
  J_1[\vec{S}] &= \frac{1}{\sqrt{n}}\sum_i^n S_i,\label{eq:J1def} \\
  J_3[\vec{S}] &= \frac{4}{3n^{\frac{3}{2}}} \sum_i^n
  \Big(S_i^3 - \frac{3}{n} |\vec{S}|^2 S_i + 3 S_i\Big). \label{eq:J3def}
\end{align}
\end{subequations}
The sum in Eq.~\eqref{eq:J1int} evaluates to a non-zero value
only if for every lattice site containing the source term $J_1$ there
exists exactly one bond terminating on that site. Similarly the
sum in Eq.~\eqref{eq:J3int} evaluates to a non-zero value only if
for the sites with branch-point $J_3$, there are exactly three bonds
terminating on that site. The factors of $1/\sqrt{n}$ in
Eq.~\eqref{eq:J1int} and $\sqrt{n}$ in Eq.~\eqref{eq:J3int} enforce
the no-loops condition necessary to obtain the statistics of a
self-avoiding polymer.  

In general, for every valid configuration with
$N_p$ connected graphs, $N_1$ end-points, and $N_3$ branch-points the
product of all the lattice site integrals gives
\begin{equation}\label{eq:C2}
  \lim_{n\to0} \prod_{x}^V \tr_\vec{S} \mathcal{C}_\alpha[\{\vec{S}\}] =
  \lim_{n\to0} n^{N_p+\frac{1}{2}(N_3-N_1)}.
\end{equation}
Using Eq.~\eqref{eq:noloop}, the exponent in Eq.~\eqref{eq:C2} is equal to
the number of loops, so in the $n\to0$ limit only those valid
configurations with no loops evaluate to 1, while all others evaluate
to 0.

It is important to note that the conditions given in
Eqs.~\eqref{eq:J1int} and \eqref{eq:J3int} do not forbid the multiple
source terms sharing the same lattice site. This oversight can be
remedied by redefining the partition function (see
Eqs.~\eqref{eq:hamil} and \eqref{eq:Z1}) such that the exponential of
the source terms is replaced by the constant and linear terms in the
power series expansion
\begin{multline}\label{eq:Z3}
  \mathcal{Z}_n(K,f_1,f_3;V) = \prod_x^V \tr_{\vec{S}_x} e^{K \!\!\!
    \sum\limits_{\langle x, x' \rangle} \!\!\! \vec{S}_x \cdot
    \vec{S}_{x'}}\\ \times \prod_x^V \big(1+f_1 J_1[\vec{S}_x]+f_3
  J_3[\vec{S}_x]\big).
\end{multline}
The structure in Eq.~\ref{eq:Z3} ensures that there will be at most
one single source term per lattice site and otherwise does not have any
impact on the derivation presented above. From here on, we will
only consider the partition function presented in Eq.~\ref{eq:Z3} for
the branched polymers system.

\subsection*{Mean Field Hamiltonian}\label{sec:MFH}
We can now convert the lattice model over a
discrete set of $\vec{S}$-vectors into a continuous field theory
$\psi(x)$ using a Hubbard-Stratonovich transformation and connect the
lattice fugacities $K$, $f_1$, and $f_3$ to physical quantities of
chemical potential and concentration in the mean field approximation.

It is necessary to carefully treat the sum over nearest neighbors on
the lattice given in Eq.~\eqref{eq:Z3} in order to change the lattice model into a continuous field theory. The sum can be written as a double sum over lattice sites
multiplied by a nearest neighbor delta function
\begin{equation}\label{dsum}
  \sum_{\langle x, x' \rangle} = \frac{1}{2} \sum_x^V \sum_{x'}^V 
  \delta_{\langle x, x' \rangle}.
\end{equation}
The function $\delta_{\langle x, x'\rangle}$ is similar to a Kronecker
delta function, and is explicitly defined as an operator that
evaluates to 1 when $x$ and $x'$ are neighbors and 0 otherwise. The
additional factor of half prevents double counting. Using
Eq.~\eqref{dsum}, we now perform a Hubbard-Stratonovich transformation
to introduce the auxiliary field $\psi(x)$
\begin{multline}\label{eq:HST}
  e^{\frac{K}{2} \sum\limits_{x,x'} \delta_{\langle x,x' \rangle}
    \vec{S}_x\cdot \vec{S}_{x'}} = \\ \int_{-\infty}^\infty
  \mathcal{D} \psi \; e^{-\frac{1}{2} \sum\limits_{x,x'}
    \delta_{\langle x, x' \rangle}^{-1} \vec{\psi}(\vec{x}) \cdot
    \vec{\psi}(\vec{x}') + \sum\limits_{x}\sqrt{K}\vec{\psi}(\vec{x})\cdot
    \vec{S}_x},
\end{multline}
where $\delta_{\langle x, x'\rangle}^{-1}$ is the inverse of the
$\delta_{\langle x, x'\rangle}$ operator. Using Eq.~\eqref{eq:HST}, the
partition function Eq.~\eqref{eq:Z3} can be written as 
\begin{multline} \label{m_integ}
  \mathcal{Z}_n(K,f_1,f_3;V) = \int_{-\infty}^\infty \mathcal{D} \psi \;
    e^{-\frac{1}{2} \sum\limits_{x,x'} \delta_{\langle x, x' \rangle}^{-1}
      \vec{\psi}(\vec{x}) \cdot \vec{\psi}(\vec{x}')}\\
  \times \prod_x^V \tr_{\vec{S}_x} e^{\sqrt{K}\vec{\psi}(\vec{x})
    \cdot \vec{S}_x}
  \big(1+f_1 J_1[\vec{S}_x]+f_3 J_3[\vec{S}_x]\big).
\end{multline}
The first term in the second line of Eq.~\eqref{m_integ} is simply
the generating function performed in Eq.~\eqref{eq:gen1}. We now use
Eqs.~\eqref{eq:J1int} and \eqref{eq:J3int} to evaluate the integrals
associated with the sources $J_1$ and $J_3$ in Eq.~\eqref{m_integ} in
the $n\to0$ limit. Without loss of generality, the source terms can pick a
special direction. To simplify the integrations, we choose the
$(1,0,\ldots,0)$ direction and thus Eqs.~\eqref{eq:J1int} and
\eqref{eq:J3int} can be written
\begin{align}\label{eq:J1int2}
  \lim_{n\to0} \tr_{\vec{S}} S_i J_1[\vec{S}] & = \delta_{i1},\\
  \lim_{n\to0} \tr_{\vec{S}} S_i S_j S_k J_3[\vec{S}] &= 
  \delta_{i1}\delta_{j1}\delta_{k1}.\label{eq:J3int2}
\end{align}
Inserting Eqs.~\eqref{eq:J1int2} and \eqref{eq:J3int2} into
Eq.~\eqref{m_integ} and performing the integral over $\vec{S}$-vector,
the partition function in the $n\to0$ limit becomes
\begin{multline}\label{eq:Z5}
  \mathcal{Z}_n(K,f_1,f_3;V) =    \int \!\! \mathcal{D} \psi \;
    e^{-\frac{1}{2} \sum\limits_{x,x'} \delta_{\langle x, x' \rangle}^{-1}
      \vec{\psi}(\vec{x}) \cdot \vec{\psi}(\vec{x}')}\\
  \times \prod_x^V 
  \bigg(1+\frac{K}{2} |\vec{\psi}(\vec{x})|^2 + f_1\sqrt{K} \psi_1(\vec{x}) +
  \frac{f_3 K^{\frac{3}{2}}}{6} \psi_1^3(\vec{x})\bigg).
\end{multline}
The $f_1 \psi_1(x)$ and $f_3
\psi_1^3(x)$ terms are proportional to the end and branch-point
densities, while $\psi^2$ is proportional to the monomer
density. Since for most physically relevant systems the ratio of
branch or end-points to monomers is low, the $f_1 \psi_1(x)$ and $f_3
\psi_1^3(x)$ terms will be much smaller than the $\psi^2(x)$ term.
Raising the second line of Eq.~\eqref{eq:Z5} into the exponent ($1+X =
e^{ln(1+X)}$) and expanding the logarithm,
we define a new effective Hamiltonian as given in Eq.~\eqref{eq:effHam}.

\section*{References}
\bibliography{branched}
\end{document}